\documentclass{PoS}
\usepackage{graphicx}

\usepackage{amssymb}
\usepackage{amsmath}
\usepackage{mathrsfs}
\usepackage{dsfont}

\usepackage{color}

\title{Lepton anomalous magnetic moments from twisted mass fermions}

\ShortTitle{Lepton anomalous magnetic moments from twisted mass fermions}

\author{Florian Burger\\
        Humboldt-Universit\"at zu Berlin, Institut f\"ur Physik, Newtonstr. 15, D-12489 Berlin, Germany\\
        E-mail: \email{florian.burger@physik.hu-berlin.de}}

\author{\speaker{Grit Hotzel}\\
        Humboldt-Universit\"at zu Berlin, Institut f\"ur Physik, Newtonstr. 15, D-12489 Berlin, Germany\\
        E-mail: \email{grit.hotzel@physik.hu-berlin.de}}

\author{Karl Jansen\\
        John von Neumann Institute for Computing (NIC), DESY, Platanenallee 6, D-15738 Zeuthen, Germany\\
        E-mail: \email{karl.jansen@desy.de}}

\author{Marcus Petschlies \\
        The Cyprus Institute, P.O. Box 27456, 1645 Nicosia, Cyprus\\
        E-mail: \email{m.petschlies@cyi.ac.cy}}

\abstract{We present our results for the leading-order hadronic quark-connected contributions to the electron, the muon, and the tau anomalous
magnetic moments obtained with four dynamical quarks. Performing the continuum limit and an analysis of systematic effects, full agreement with
phenomenological results is found. To estimate the impact of omitting the quark-disconnected contributions to the hadronic vacuum polarisation we
investigate them on one of the four-flavour ensembles. Additionally, the light quark contributions on the four-flavour sea are compared to the values
obtained for $N_f=2$
physically light quarks. In the latter case different methods to fit the hadronic vacuum polarisation function are tested.}

\FullConference{The 32nd International Symposium on Lattice Field Theory\\
		 23-28 June, 2014\\
		 Columbia University New York, NY}

\begin{document}

\section{Introduction}

The hadronic vacuum polarisation function governs the leading hadronic contributions of several electroweak parameters~\cite{Renner:2012fa}.
In particular, we can extract from it the hadronic leading-order anomalous magnetic moments of all three leptons present in the standard model of
particle interactions.
In this proceedings contribution we compare their values with the phenomenological results obtained when using a dispersion relation.
Additionally, we confirm the results of our earlier chiral
extrapolations by computing the light-quark contributions at the physical value of the pion mass.
Furthermore, our attempts to reduce remaining systematic uncertainties are described. The status of investigations of
disonnected
contributions and different fitting strategies are shown.

\section{Basic equations}
The leading-order hadronic contribution to the lepton anomalous magnetic moments in Euclidean space-time is given by~\cite{Blum:2002ii}
\begin{equation}
a_{\mathrm{l}}^{\mathrm{hvp}} = 
\alpha^2 \int_0^{\infty} \frac{d Q^2 }{Q^2} 
w\left( \frac{Q^2}{m_{\mathrm{l}}^2}\right) \Pi_{\mathrm{R}}(Q^2) \; ,
\label{eq:amudef}
\end{equation}
where $\alpha$ is the fine structure constant, $Q^2$ the Euclidean momentum, $m_{\mathrm{l}}$ the lepton mass, and
$\Pi_{\mathrm{R}}(Q^2)$ the
renormalised hadronic vacuum polarisation function,
%\begin{equation}
$\Pi_{\mathrm{R}}(Q^2)= \Pi(Q^2)- \Pi(0) $ ,
%\label{eq:pirenorm}
%\end{equation}
obtained from the vacuum polarisation tensor 
\begin{equation}
   \Pi_{\mu \nu}(Q)= \int d^4 x \,e^{iQ\cdot(x-y)} \langle J_{\mu}(x) J_{\nu}(y)\rangle = (Q_{\mu} Q_{\nu} - Q^2
\delta_{\mu
\nu}) \Pi(Q^2)\; ,
\label{eq:vptensor}
\end{equation}
which is the correlator of two electromagnetic vector currents $J_\mu(x)$. In the lattice computations of the quark-connected diagrams contributing to
$a_{\mathrm{l}}^{\mathrm{hvp}}$ we employ the conserved point-split vector current. The weight function $w\left( Q^2/ m_{\mathrm{l}}^2\right)$ is
known and becomes maximal at $Q^2_{\rm max}=(\sqrt{5}-2)m_l^2$.

Since the computations with $N_f=2+1+1$ quarks have been performed at unphysically large pion masses, a chiral extrapolation to the physical point is mandatory.
To simplify this extrapolation, we use in the four-flavour case the same redefinition as in~\cite{Feng:2011zk, Renner:2012fa, Burger:2013jya}
\begin{equation}
 a_{\overline{\mathrm{l}}}^{\mathrm{hvp}} = \alpha^2 \int_0^{\infty} \frac{d Q^2 }{Q^2} w\left( \frac{Q^2}{H^2}
\frac{H_{\mathrm{phys}}^2}{m_{\mathrm{l}}^2}\right) \Pi_{\mathrm{R}}(Q^2) 
\label{eq:redef}
\end{equation}
with the hadronic scale $H=m_V$, the lowest lying vector meson state. $H=H_{\rm phys}=1$ corresponds to the standard definition in
Eq.~(\ref{eq:amudef}).

The hadronic vacuum polarisation function defined as in~\cite{Burger:2013jya} is fitted by dividing the momentum range between $0$ and $100\,{\rm
GeV}^2$ in a low-momentum region $ 0 \le Q^2 \le 2\,{\rm GeV}^2$ and a high-momentum one $ 2\,{\rm GeV}^2 < Q^2 \le 100\,{\rm GeV}^2$ according to
\begin{equation}
 \Pi(Q^2) = (1- \Theta(Q^2-Q^2_{\rm match}))\Pi_{\mathrm{low}}(Q^2) + \Theta(Q^2-Q^2_{\rm match}) \Pi_{\mathrm{high}}(Q^2) \; ,
\label{eq:pilowandhigh}
\end{equation}
where the low-momentum fit function is given by
\begin{equation}
  \Pi_{\mathrm{low}}(Q^2) = \sum_{i=1}^M \frac{f^2_i}{m^2_i + Q^2} + \sum_{j=0}^{N-1} a_j (Q^2)^{j} \; ,
\label{eq:pilow}
\end{equation}
and the form of the high-momentum part is inspired by perturbation theory
\begin{equation}
  \Pi_{\mathrm{high}}(Q^2) = \log(Q^2) \sum_{k=0}^{B-1} b_k (Q^2)^{k}  + \sum_{l=0}^{C-1} c_l (Q^2)^{l} \; . 
\label{eq:pihigh}
\end{equation}
This defines our so-called MNBC fit function, e.g. M1N2B4C1 means $M=1$, $N=2$, $B=4$, and $C=1$ in Eqs.~(\ref{eq:pilow}) and (\ref{eq:pihigh})
above. 
% The value of $Q^2_{\rm match}$ in the Heaviside functions in Eq.~(\ref{eq:pilowandhigh}) is $2\,{\rm GeV}^2$. Varying it between $1\,{\rm
% GeV}^2$ and $3\,{\rm GeV}^2$ does not lead to observable differences as long as the transition between the low- and the high-momentum part of the fit
% is smooth.

\section{$N_f=2+1+1$ fermions at unphysical pion masses}

Since the momentum, where the weight function appearing in the definition of $a_{\mathrm{l}}^{\rm hvp}$ in
Eq.~(\ref{eq:amudef}) attains its maximum, is proportional to the squared lepton mass and the lepton masses vary over four orders of magnitude, the
different lepton
anomalous magnetic moments are sensitive to very different momentum regions. In addition, the electron's magnetic moment is one of the experimentally
and theoretically most precisely known physical quantities and thus provides a meaningful cross-check of the method used to compute the muon anomalous
magnetic moment for which a well-known discrepancy exists. 

\subsection{Quark-connected contributions}
\begin{figure}[htb]
\centering
 \includegraphics[width=0.65\textwidth]{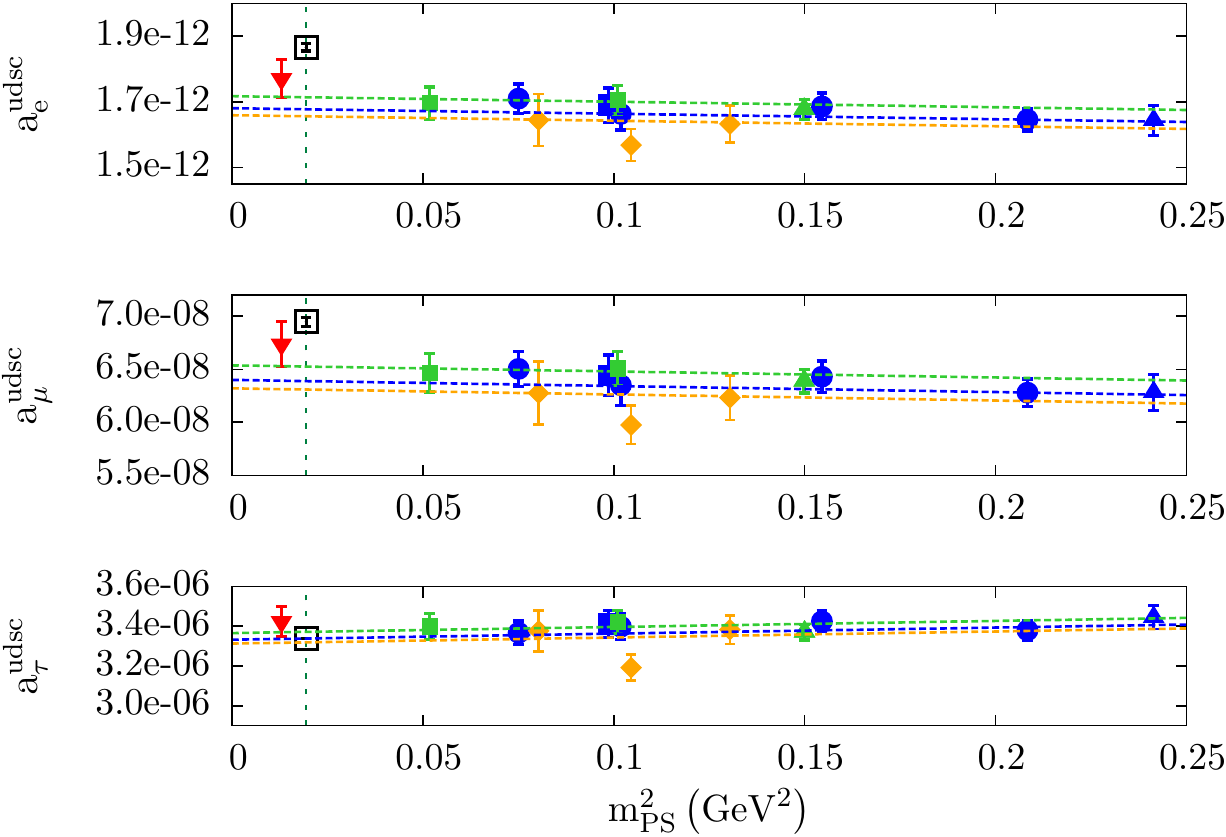}
\caption{Chiral and continuum extrapolation of the connected leading hadronic contributions to the three lepton anomalous magnetic moments. The
combined extrapolation has been performed with
$a_{\mathrm{lepton}}(m_{\rm PS}, a) = A + B~m_{PS}^2 + C~a^2$. The dotted orange line shows $a_{\mathrm{lepton}}(m_{\rm PS}, 0.086\,{\rm fm})$, the
blue line corresponds to $a_{\mathrm{lepton}}(m_{\rm PS}, 0.078\,{\rm fm})$, and the green line depicts  $a_{\mathrm{lepton}}(m_{\rm PS}, 0.061\,{\rm
fm})$. The inverted red triangle shows the value in the continuum limit at the physical value of the pion mass. It has been displaced to the left to
facilitate the comparison with the dispersive results in the black squares for $a_{\mathrm{e}}^{\rm hvp}$~\cite{Nomura:2012sb},
$a_{\mathrm{\mu}}^{\mathrm{hvp}}$~\cite{Hagiwara:2011af}, and $a_{\mathrm{\tau}}^{\mathrm{hvp}}$~\cite{Eidelman:2007sb}.} 
\label{fig:gm2_all}
\end{figure}

Conducting exactly the same analysis as described in~\cite{Burger:2013jya} for the anomalous magnetic moment of the muon, only changing the lepton
masses in the numerical integration, we have computed the leading hadronic contributions to the anomalous magnetic moments of the electron and the
$\tau$-lepton. The results and their chiral and continuum extrapolation are depicted in Fig.~\ref{fig:gm2_all}. For completeness we also show the
muon anomalous magnetic moment. 

Our four-flavour
results extrapolated to the physical pion mass in the continuum limit can directly be compared  with phenomenological results relying on a
dispersion relation.
%  connecting the
% hadronic vacuum polarisation to experimental $e^+ e^-$-scattering or $\tau$-decay data. 
This is shown in Tab.~\ref{tab:results_gm2}. Here, we have
also included estimates of the systematic uncertainties of our results which have been obtained exactly in the same way as in~\cite{Burger:2013jya}.
As for the muon, the only non-negligible systematic uncertainties arise from excited state contaminations in the fit of the vector meson parameters
and the number of terms in our MNBC fit functions in Eqs.~(\ref{eq:pilow}) and (\ref{eq:pihigh}). However, we have not yet satisfactorily quantified
the systematic uncertainty
from neglecting the quark-disconnected contributions.

\small{
\begin{table}[htb]
\begin{center}
\begin{tabular}{c|c c c}
 &  $a_{\mathrm{e}}^{\rm hvp}$ & $a_{\mathrm{\mu}}^{\mathrm{hvp}}$ & $a_{\mathrm{\tau}}^{\mathrm{hvp}}$ \\
\hline \hline
 & & & \vspace{-0.40cm} \\
this work & $1.77(06)(05)\cdot 10^{-12}$ & $6.74(21)(18)\cdot 10^{-8}$ &   $3.42(08)(05)\cdot 10^{-6}$ \\
dispersive analyses & $1.87(01)(01)\cdot 10^{-12}$~\cite{Nomura:2012sb} & $6.95(04)(02) \cdot 10^{-8}$~\cite{Hagiwara:2011af}& 
$3.38(04)\cdot 10^{-6}$~\cite{Eidelman:2007sb}\\
\end{tabular}
\caption{\label{tab:results_gm2} Comparison of our first-principle values for $a_{\mathrm{e}}^{\rm hvp}$, $a_{\mathrm{\mu}}^{\mathrm{hvp}}$, and
$a_{\mathrm{\tau}}^{\mathrm{hvp}}$ with phenomenological results.} 
\end{center}
\end{table}}

Taking statistical as well as systematic uncertainties into account, full agreement is found between our lattice results for the quark-connected
contributions and the phenomenological continuum values for all three leptons. As mentioned before, this constitutes a non-trivial cross-check of our
computation of the leading hadronic contribution to the muon $(g-2)$. 
% Furthermore, this provides an indication that the quark-disconnected
% diagrams contribute only very little to the $a_{\mathrm{l}}^{\rm hvp}$. In the next subsection we want to scrutinise this statement. 

Another important cross-check is provided by comparisons with results obtained from different fermion discretisations. Since the HPQCD
collaboration recently provided very precise values for the strange and charm quark contributions to
$a_{\mathrm{\mu}}^{\mathrm{hvp}}$~\cite{Chakraborty:2014mwa} obtained from a dedicated effort, we state our results computed in~\cite{Burger:2013jya} in the continuum limit:
\begin{eqnarray}
a_{\mathrm{\mu},\mathrm{s}}^{\mathrm{hvp}} & = & 5.36(19)\cdot 10^{-9} \\
a_{\mathrm{\mu},\mathrm{c}}^{\mathrm{hvp}} & = & 1.418(61)\cdot 10^{-8} \; .
\end{eqnarray}
They are compatible with the values obtained by HPQCD, but have larger statistical uncertainties, since they originate from only approximately 150
configurations per ensemble as this already gave smaller uncertainties than obtained for the light quark contributions.

\subsection{Quark-disconnected contributions}
Quark-disconnected Feynman diagrams naturally arise when Wick contracting the fields in the current correlator. In most existing lattice calculations
of the leading hadronic contributions to lepton anomalous magnetic moments they have been neglected due to their large computational cost.

\begin{figure}[htb]
\centering
\includegraphics[width=0.6\textwidth]{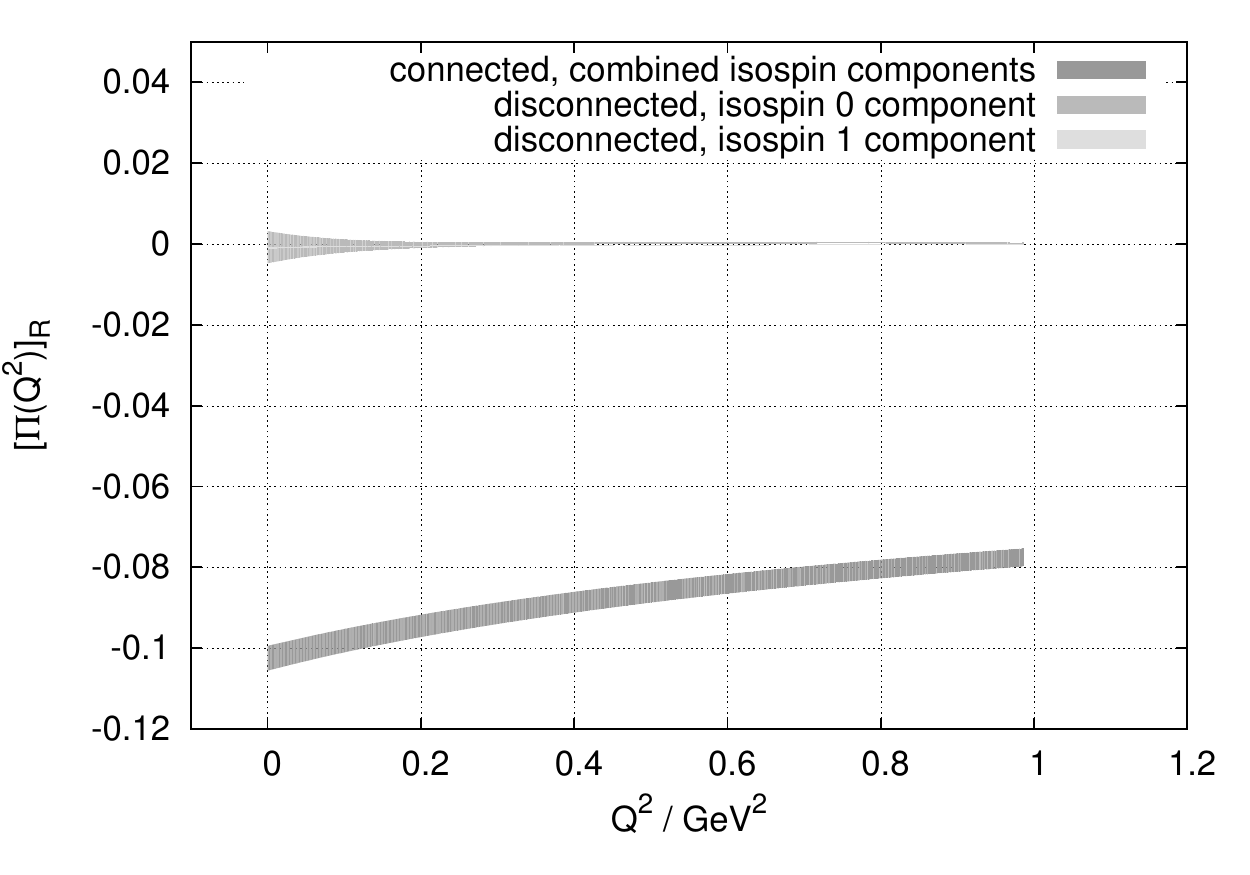}
\caption{Comparison of the light quark contributions to the unsubtracted hadronic vacuum polarisation function from quark-connected and disconnected
diagrams of the local current correlator. The subscript $\rm R$ signals that the renormalisation factor $Z_V$ is included. The values have
been obtained with the analytical continuation method described in~\cite{Feng:2013xsa} without correcting for finite-size effects.} 
\label{fig:disconnected}
\end{figure}

In order to remedy this shortcoming we have started investigating the disconnected contributions on one of our $N_f=2+1+1$ ensembles, namely B55.32
(see~\cite{Baron:2010bv, Baron:2010th} for details), featuring $m_{\rm
PS} \approx 390\, {\rm
MeV}$ and $a\approx 0.08\,{\rm fm}$. However, for the
point-split vector current we have not, yet, succeeded to detect a signal. Only when analysing the current correlator of two local
vector currents, we observe a signal for the light quark contribution. The resulting contribution to the hadronic vacuum polarisation is
significantly smaller than the connected contribution as can be
seen in Fig.~\ref{fig:disconnected}. Here, we have used 24 stochastic volume sources on 1548 configurations and 48 stochastic volume sources on 4996
configurations for both isospin components.  
The necessary renormalisation factor $Z_V$ has been
obtained from the ratio of the connected conserved and local current correlators. 

Despite the rather large statistics of more than 6500 configurations, the isoscalar components of the anomalous magnetic moments of all leptons are
zero within the error and might even be negative as predicted in~\cite{DellaMorte:2010aq}.
However, when using analytical continuation~\cite{Feng:2013xsa} their signs change when changing the maximal time slice up to which the current
correlator is summed. Furthermore, the size of the uncertainty grows with the number of time slices due to essentially only adding up noise after
time slice two.

\section{$N_f=2$ fermions at the physical point}

\subsection{Light quark contributions to lepton anomalous magnetic moments}
\begin{figure}[htb]
\centering
\includegraphics[width=0.6\textwidth]{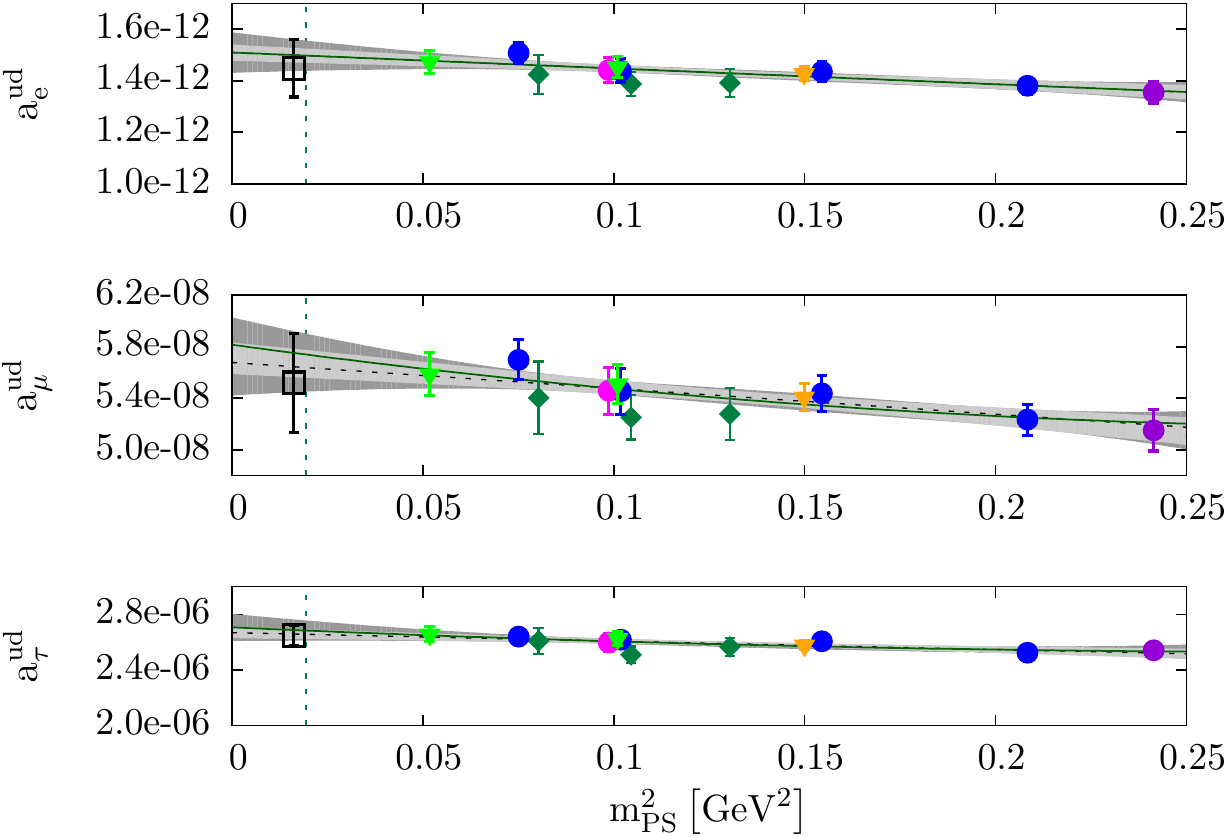}
\caption{Comparison of the chiral extrapolation of the light quark contributions to the three lepton anomalous magnetic moments with the values 
obtained with the standard definition Eq.~(\protect\ref{eq:amudef}) at the physical value of the pion mass (black square).} 
\label{fig:gm2_light_all}
\end{figure}

When determining $a_{\rm{l}}^{\rm hvp}$ on the ETMC's $N_f=2+1+1$ ensembles~\cite{Baron:2010bv, Baron:2010th} one potential source of a systematic
error is the chiral extrapolation to the physical pion mass. Meanwhile an ensemble with $N_f=2$ dynamical
quarks at the physical point~\cite{Abdel-Rehim:2013yaa} has been generated. We have computed the light quark contributions to the lepton anomalous
magnetic moments with the standard
definition Eq.~(\ref{eq:amudef}) on 804 configurations and found full agreement with our old extrapolated $N_f=2$ as well as $N_f=2+1+1$
results. The extrapolations are
depicted in Fig.~\ref{fig:gm2_light_all} whereas the numbers also including the old $N_f=2$ values from~\cite{Feng:2011zk} are given in
Tab.~\ref{tab:results_gm2_light}. In contrast to~\cite{Burger:2013jya} we have employed M1N3B4C1 fits, i.e.~one additional fit parameter, due to
having more than three times the statistics than for the $N_f=2+1+1$ ensembles.

\small{
\begin{table}[htb]
\begin{center}
\begin{tabular}{c|c c c}
 &  physical point & extrapolated $N_f=2$ & extrapolated $N_f=2+1+1$ \\
\hline \hline
 & & & \vspace{-0.40cm} \\
$a_{\rm e }^{\rm hvp}$ & $1.45(11)\cdot 10^{-12}$ & $1.51(04)\cdot 10^{-12}$ & $1.50(03)\cdot 10^{-12}$  \\
$a_{\rm \mu}^{\rm hvp}$ & $5.52(39)\cdot 10^{-8}$ & $5.72(16)\cdot 10^{-8}$ & $5.67(11)\cdot 10^{-8}$\\
$a_{\rm \tau}^{\rm hvp}$ & $2.65(07)\cdot 10^{-6}$ & $2.65(02)\cdot 10^{-6}$ & $2.66(02)\cdot 10^{-6}$ \\
\end{tabular}
\caption{\label{tab:results_gm2_light} Comparison of the values for $a_{\mathrm{e}}^{\rm hvp}$, $a_{\mathrm{\mu}}^{\mathrm{hvp}}$, and
$a_{\mathrm{\tau}}^{\mathrm{hvp}}$ obtained at the physical point using the standard definition Eq.~(\protect\ref{eq:amudef}) with the
results of the linear extrapolations from our improved definition Eq.~(\protect\ref{eq:redef}) on the old $N_f=2$
and $N_f=2+1+1$ ETMC ensembles.} 
\end{center}
\end{table}}

\subsection{Different fit functions}
In~\cite{Aubin:2012me} the authors suggested for the first time that employing Pad\'e fits to parametrise the hadronic vacuum polarisation function
yields a completely model-independent determination of $a_{\rm{\mu}}^{\rm hvp}$. In the following we will use the notation of this older paper and
compare the results of [0,1] and [1,1] Pad\'e fits with our M1N2 and M1N3 fits up to $Q_{\rm max}^2=0.75\,{\rm GeV}^2$. This already
stretches the applicability of Pad\'e fits which only seem to describe the data well at small momenta. Thus, we limit the
comparison to the case of the electron where the weight function guarantees an early saturation of the integral.

\small{
\begin{table}[htb]
 \begin{center}
 \begin{tabular}{c | c  | c  |  c  | c }
 
 & M1N2 & M1N3 (standard) & [0,1] Pad\'e fit & [1,1] Pad\'e fit \\
 \hline \hline
 $a^2 \times$ pole &  0.154(38) &  0.154(38) & 0.183(01)  & 0.188(02)  \\
 $a_{\rm e}^{\rm hvp}$ & $1.45(11)\cdot 10^{-12} $ & $1.56(09)\cdot 10^{-12}$ &   $1.31(05)\cdot 10^{-12}$&  $1.67(20)\cdot 10^{-12}$
\end{tabular}
\caption{\label{tab:fitfunctions} Comparison of the single pole and the value for $a_{\rm e}^{\rm hvp}$ obtained from MN and Pad\'e fits, both with
standard
definition Eq.~(\protect\ref{eq:amudef}).}
 \end{center}
\end{table}}

We observe first of all that the values for the fitted single pole obtained from MN fits and Pad\'e fits are compatible. Secondly, the results for
$a_{\rm e}^{\rm hvp}$ from the [0,1] Pad\'e and the M1N2
fit (3 free parameters) as well as those from the [1,1] Pad\'e fit and the M1N3 fit (4 free parameters) are mutually consistent. This is no surprise
as
the Pad\'e fits and the MN fits with the same number of parameters only differ by lattice artefacts, since in the
MN fits the pole is determined from the temporal correlator whereas in the Pad\'e fits the pole comes from $\Pi(Q^2)$. We thus
expect equivalence in the continuum limit provided the same procedure is followed in both cases (same number of parameters, standard definition for $a_{\rm l}^{\rm hvp}$, keeping correlations and properly propagating uncertainties from vector meson fits).

\section{Conclusions}
In this proceedings contribution we have computed the leading hadronic quark-connected contributions to the anomalous magnetic moments of all three
leptons on ensembles featuring $N_f=2+1+1$ twisted mass fermions and found full agreement with their phenomenological values. Furthermore, we have
for the first time reported a signal for the disconnected contributions on one of those ensembles. In order to check the values obtained for the
light quark contributions from using the improved definition of $a_{\rm{l}}^{\rm hvp}$, we have performed a computation at the physical value of the
pion mass on a $N_f=2$ ensemble. This fully confirms our earlier results. We have also investigated different fit functions and are studying the
all-mode-averaging technique of~\cite{Blum:2012uh}.

\section*{Acknowledgements}
Special thanks goes to the authors of~\cite{Michael:2013gka} who generously granted us access to their data for the disconnected contributions of the
local vector current correlators.
This work has been supported in part by the DFG Corroborative
Research Center SFB/TR9.
G.H.~gratefully acknowledges the support of the German Academic National Foundation (Studienstiftung des deutschen Volkes e.V.)
 and of the DFG-funded Graduate School GK 1504.
K.J. was supported in part by the Cyprus Research Promotion
Foundation under contract $\Pi$PO$\Sigma$E$\Lambda$KY$\Sigma$H/EM$\Pi$EIPO$\Sigma$/0311/16.
The numerical computations have been performed on the {\it SGI system HLRN-II} and the
{\it Cray XC30 system HLRN-III} at the {HLRN Supercomputing Service Berlin-Hannover},  FZJ/GCS, BG/P, and BG/Q at FZ-J\"ulich.

\end{document}